\documentclass[aps,prd,nobibnotes,twocolumn,superscriptaddress,bibliography]{revtex4-2}

\usepackage{amsfonts}
\usepackage{mathrsfs}
\usepackage{amsmath}
\usepackage{color}
\usepackage{graphicx}
\usepackage{bm}
\usepackage{amssymb}

\usepackage[colorlinks=true, letterpaper=true, pdfstartview=FitV, linkcolor=blue, citecolor=blue, urlcolor=blue]{hyperref}
\usepackage[hyphenbreaks]{breakurl}

\makeatother
\begin{document}
	\title{Balancing chemical equations: form the perspective of Hilbert basis}

\author{Zeying Zhang}\email{zzy@mail.buct.edu.cn}
\affiliation{College of Mathematics and Physics, Beijing University of Chemical Technology, Beijing 100029, China}
	
\author{Xueqin Zhang}
\affiliation{School of Light Industry Science and Engineering, Beijing Technology and Business University, Beijing 100048, China}

\author{Y. X. Zhao}
\affiliation{Department of Physics and HKU-UCAS Joint Institute for Theoretical	and Computational Physics at Hong Kong, The University of Hong Kong,
Pokfulam Road, Hong Kong, China}

\author{Shengyuan A. Yang}
\affiliation{Research Laboratory for Quantum Materials, IAPME, FST, University of Macau, Macau SAR, China}

\begin{abstract}
The balancing of chemical equations is a basic problem in chemistry.
A commonly employed method is to convert the task to a linear algebra problem, and then solve the null space of the constructed formula matrix. However, in this method, the directly obtained solution may be invalid, and there is no canonical choice of independent basis reactions. Here, we show that these drawbacks originate from the fact that the fundamental structure of solutions here is not a linear space but a positive affine monoid.
This new understanding enables a systematic approach and a complete description of all possible reactions by a unique set of independent elementary reactions, called Hilbert-basis reactions.
By clarifying its underlying mathematical structure, our work offers a new perspective on this old problem of balancing chemical equations.

\end{abstract}
\maketitle

\section{Introduction}

For a given set of $m$ reactants $\{R_i\}$ $(i=1,\cdots,m)$ and a given set of $n$ products $\{P_j\}$ $(j=1,\cdots,n)$,
balancing chemical equations for them refers to finding nonnegative integers $c_k\geq 0$ $(k=1,2,\cdots,m+n)$ such that
\begin{equation}\label{CE}
  c_1 R_1 +\cdots+c_m R_m\rightarrow c_{m+1}P_1 +\cdots c_{m+n}P_n
\end{equation}
is a valid chemical equation. This is a very basic problem in chemistry.
For simple cases, one may obtain the solution just by inspection \cite{toth_balancing_1997}. For more complicated cases, it is difficult to solve using just inspection. One needs to resort to certain techniques, e.g., by using oxidation numbers or half reactions \cite{waldbauer_notes_1926, simons_chemical_1926, jette_balancing_1927, bennett_balancing_1935, morris_balancing_1938,pernicone_method_1981, garcia_new_1987,herndon_balancing_1997, zheng_balancing_2022, yuen_h-atom_2022, yuen_balancing_2022}, which requires a fair knowledge of chemistry.

There also exist a class of approaches known as algebraic methods, which translate the task into an algebraic problem and solve it by mathematical techniques. Such algebraic methods were initially attempted by Bottemly more than a century ago \cite{bottomley_note_1878}, and further  developed by many others over the years \cite{scheflan_balancing_1932, deming_balancing_1934, krishnamurthy_generalized_1978, carrano_balancing_1978, smith_what_1979, alberty_balancing_1983, alberty_conversion_1992,wink_use_1994, olson_analysis_1997,risteski_new_2009,risteski_new_2014}. For most of them, the procedures can be formulated in terms of matrices, and the key  is essentially to find the null space $\text{ker}\mathbf F$ of a matrix $\mathbf F$ known as the formula matrix. Hence, algebraic methods are especially suitable for computer solutions \cite{missen_question_1989, weltin_let_1994,smith_using_1997, luque_ruiz_design_2000, kumar_computer_2001, sen_chemical_2006, candelario-aplaon_notitle_2019,altintas_application_2022}.

Nevertheless, there are in general two shortcomings of the existing algebraic methods. First, as a linear space, it is clear that the null space $\text{ker}\mathbf F$ must also contain invalid solutions which do not fulfill the requirement that all components $c_k$ are nonnegative integers. Denoting the set of all valid solutions by $S$, this means $S$ is only a proper subset of $\text{ker}\mathbf F$. Second, the solution $\text{ker}\mathbf F$ is usually represented by its basis vectors, but there is no canonical (unique) choice of the bases for a linear space. One may impose certain additional conditions to select a set of bases \cite{risteski_new_2009}. However, there is generally no guarantee that all members of this set belong to $S$.

In this work, we point out that the algebraic structure of the solution set $S$ is not described by a linear space; instead, it corresponds to a structure known as positive affine monoid. The above mentioned shortcomings fundamentally originates from the inadequate treatment of $S$ as a linear space, and they are completely eliminated by using approaches for positive affine monoid. Particularly, based on this new perspective, we show that for each given system of reactants and products, there exists a \emph{unique} set of independent elementary reactions, called the Hilbert-basis reactions (HBRs), corresponding to the unique Hilbert basis of $S$. All other reactions in $S$ can be decomposed into HBRs. Our proposed approach to find HBRs can be readily performed by using existing routines. We develop a software package which implements this approach and can be readily used  on a personal computer. This work clarifies the underlying mathematical structure involved in balancing chemical equations, and offers a new approach to obtain elementary chemical reactions which avoids the previous shortcomings.

\section{algebraic structure of solution space}

Balancing chemical equations is based on the principle of conservation of atoms/charges for chemical reactions.
Let $\{E_\ell\}$ ($\ell=1,\cdots,s$) be the set of all the elements involved in the substances $\{R_i\}$ and $\{P_j\}$.
For each $E_\ell$, we use $r_{\ell i}$ ($p_{\ell i}$) to denote the number of $E_\ell$ atoms in a $R_i$ ($P_j$) molecule.
The conservation of atoms tells us for each $\ell$, equation (\ref{CE}) leads to the following equation:
\begin{equation}\label{iE}
 r_{\ell 1} c_1 +\cdots +r_{\ell m}c_m- p_{\ell 1}c_{m+1}-\cdots -p_{\ell n}c_{m+n}=0.
\end{equation}
This gives $\ell$ coupled equations.

To put these equations in a more compact form, we define $\mathbf R=[r_{\ell i}]$ as an $s\times m$ matrix, and $\mathbf P=[p_{\ell j}]$ as an $s\times n$ matrix.
Then, we construct the formula matrix as
\begin{equation}
  \mathbf F=[\mathbf R, -\mathbf P],
\end{equation}
which is an $s\times (m+n)$ matrix. One finds that Eq.~(\ref{iE}) can be put into a matrix equation
\begin{equation}\label{Fc}
  \mathbf{Fc}=\mathbf 0,
\end{equation}
where the column vector $\mathbf c=(c_1,\cdots, c_{m+n})^T$ is the solution of this equation.

This matrix equation (\ref{Fc}) is the common starting point for existing algebraic methods. It should be noted that
in our definition of formula matrix, there is a minus sign in front of the product matrix $\mathbf P$. This differs from
definition in the early works by several authors \cite{blakley_chemical_1982, alberty_balancing_1983, alberty_chemical_1991, alberty_conversion_1992,  ugi_algebraic_1992}. Because in those treatments, the solution space is taken as a linear space, so the minus sign becomes insignificant. In addition, if the substances $\{R_i\}$ and $\{P_j\}$ are not all charge neutral, the conservation of charge will give an additional equation:
\begin{equation}
 r_{0 1} c_1 +\cdots +r_{0 m}c_m- p_{0 1}c_{m+1}-\cdots -p_{0 n}c_{m+n}=0,
\end{equation}
where $r_{0i}$ ($p_{0j}$) is the charge of $R_i$ ($P_j$). One can easily see that this just adds a row into the defined formula matrix, without causing additional complexity.

In previous treatments, one solves Eq.~(\ref{Fc}) by regarding the solution space as a linear space (over rational numbers $\mathbb{Q}$). Then the task is reduced simply to find the null space $\text{ker}\mathbf F$ for matrix $\mathbf F$, which is
a routine linear algebra problem. However, as we mentioned, $\text{ker}\mathbf F$ is actually larger than the
true solution space $S$, which requires components of solution $\mathbf c$ must be nonnegative integers, i.e.,
\begin{equation}\label{S}
  S=\big\{\mathbf c\ | \ \mathbf{Fc}=0\ \text{and}\ c_k\in \mathbb{Z}_{\geq 0}, k=1,\cdots, m+n\big\}.
\end{equation}
Then, the question is: How can we describe the algebraic structure of $S$ in a more accurate way?

According to a mathematical result known as Gordan's lemma, the structure of $S$ defined in Eq.~(\ref{S}) corresponds to a positive affine monoid \cite{gubeladze_polytopes_2009}. A commutative monoid is a set with a binary operation, which is closed, associative, commutative, and has identity element. For the solution space $S$ here, its elements are the vectors with components being nonnegative integers.
The binary operation here is simply the vector addition. Clearly, if $\mathbf{a,b}\in S$, then $\mathbf a+\mathbf b$ is also an element of $S$, so it is closed. The associativity and commutativity are evident. The identity element is given by the zero vector. It is important to note that different from a group, the elements here do not have inverses, since if $\mathbf c\in S$, then $-\mathbf c\notin S$. The only exception is the zero vector, which is its own inverse, and the adjective ``positive'' refers to this feature. Finally, the word ``affine'' means that the monoid is finitely generated, i.e., it has a finite number of generators. Moreover, among the generators, there exists a unique minimal set, which is crucial for the theory of positive affine monoid, as we discuss below.

\section{Hilbert-basis reactions}

A key property of a positive affine monoid is that it has a \emph{unique minimal} set of generators, known as Hilbert basis. For monoid $S$, its Hilbert basis may be denoted as $\text{Hilb}(S)$ \cite{gubeladze_polytopes_2009}. Each element in Hilbert basis cannot be decomposed into sum of other elements in $S$. Meanwhile, every element of $S$ can be generated by elements in $\text{Hilb}(S)$, although the decomposition may not be unique. This is in contrast to the properties of linear spaces, whose choice of bases is not unique but the decomposition of a vector into given bases is unique.  Suppose $\text{Hilb}(S)=\{\mathbf v^i\}$ ($i=1,\cdots,d$) containing $d$ elements.
We may write the following expression
\begin{equation}
  S=\big\{q_1 \mathbf v^1+\cdots +q_d\mathbf v^d\ | \  q_i\in \mathbb{Z}_{\geq 0}, \mathbf v^i\in \text{Hilb}(S)\big\}.
\end{equation}
In general, the number of elements in Hilbert basis $d$ will be larger than the dimension of $\text{ker}\mathbf F$.

This set of Hilbert basis are of central importance, as they represent the basic building blocks and contain the whole information of the solution space $S$. It follows that the chemical reactions corresponding to the Hilbert basis
are also of key interest, as they correspond to the most elementary ones. They are named as HBRs. Thus, in our approach, the task of balancing chemical equations for given reactants and products is reduced to finding the corresponding HBRs.

Mathematicians have developed several algorithms for calculating the Hilbert basis for a given affine monoid \cite{cohen_mathematical_2002}. The basic idea involves two steps.
The first step is to find a set of (generally not minimal) generators of the affine monoid. This step can be done by finding the generators of rational cones, of which the intersection forms the affine monoid.
The second step is to reduce this set of generators to Hilbert basis \cite{gubeladze_polytopes_2009}. Such algorithms have been efficiently implemented in several open source packages \cite{bruns_normaliz_2010, 4ti2_team_4ti2software_nodate, the_latte_team_latte_nodate}.
In the examples below, we shall use the \textsf{Normaliz} package~\cite{bruns_normaliz_2010} for the step of obtaining the Hilbert basis.

\section{Examples}
Here, we demonstrate the application of our approach to two examples.  More examples can be found in this website: \url{ https://github.com/zhangzeyingvv/HilbertBalance/examples.nb}. We have developed a software package called \textsf{HilbertBalance} \cite{noauthor_hilbertbalance_nodate}, which implements our Hilbert basis approach to balance chemical equations on a personal computer and is made accessible as open-source on this website.

For the first example, we consider the reaction of the chlorate ion in hydrochloric acid \cite{toby_ambiguities_1994}.
The reactants consist of $R=\{\text{Cl}\text{O}_{3}^-, \text{Cl}^-, \text{H}^+\}$ and the products are given by $P=\{
\text{Cl}\text{O}_{2}, \text{Cl}_{2}, \text{H}_{2}\text{O}\}$. Hence, we need to solve $6\times 1$ vector $\mathbf c$ for the following chemical equation:
\[
	c_1\text{Cl}\text{O}_{3}^- +c_2 \text{Cl}^- +c_3 \text{H}^+\rightarrow c_4\text{Cl}\text{O}_{2}+c_5\text{Cl}_{2}+c_6\text{H}_{2}\text{O},
\]
with its components $c_i$ being nonnegative integers.

For this problem, the $\mathbf{F}$ matrix  is given by
\begin{equation}
\mathbf F=\left[
\begin{array}{crrrrrr}
\text{charge}&	-1 & -1 & 1 & 0 & 0 & 0 \\
\text{Cl}&	1 & 1 & 0 & -1 & -2 & 0 \\
\text{H}&	0 & 0 & 1 & 0 & 0 & -2 \\
\text{O}&	3 & 0 & 0 & -2 & 0 & -1 \\
\end{array}
\right],
\end{equation}
where we add the charge conservation condition in the first row of the matrix.

If one pursues the conventional approach by treating the solution set $S$ as a linear space and directly calcuates
the null space of $\mathbf F$, then the obtained $\ker\mathbf F$ is two dimensional and the basis vectors $\mathbf b_1, \mathbf b_2$ obtained from Gaussian elimination \cite{artin_algebra_2011} are
\begin{equation}
	\left\{
	\begin{array}{lrrrrrr}
 \mathbf b^1& 5 & 1 & 6 & 6 & 0 & 3 \\
\mathbf b^2 &-4 & 4 & 0 & -6 & 3 & 0 \\
	\end{array}
	\right\}.
\end{equation}
Here, for compactness, we express the column vector basis as rows inside the brackets. One can easily spot the problem here: the basis $\mathbf b_2$ is not a valid solution since it contains negative coefficients. One may get rid of the negative coefficients by making linear combinations of the two vectors, but there is no guarantee that a general linear combination would be a valid solution. The choice of basis here is not unique. More importantly, although the dimension of linear space
$\ker\mathbf F$ is two,
this does not correspond to the number of independent basis reactions.

Now, let's apply our new approach by looking for the Hilbert basis for the solution set $S$. We find that
this determines a unique set of three Hilbert basis
\begin{equation}
\left\{
\begin{array}{lcccccc}
&\text{Cl}\text{O}_{3}^-&\text{Cl}^-&\text{H}^+&\text{Cl}\text{O}_{2}&\text{Cl}_{2}&\text{H}_{2}\text{O}\\
\mathbf v^1 &1 & 5 & 6 & 0 & 3 & 3 \\
\mathbf v^2&2 & 2 & 4 & 2 & 1 & 2 \\
\mathbf v^3 &5 & 1 & 6 & 6 & 0 & 3 \\
\end{array}
\right\}.
\end{equation}
Here, for better understanding, in the top arrow, we indicate the substance for each entry of the vector.
This shows there are three HBRs:
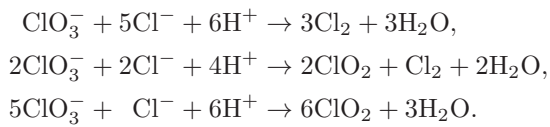
\begin{align*}
\text{Cl}\text{O}_{3}^-+5\text{Cl}^-+6\text{H}^+&\rightarrow 3\text{Cl}_{2}+3\text{H}_{2}\text{O},\\
2\text{Cl}\text{O}_{3}^-+2\text{Cl}^-+4\text{H}^+& \rightarrow 2\text{Cl}\text{O}_{2}+\text{Cl}_{2}+2\text{H}_{2}\text{O},\\
5\text{Cl}\text{O}_{3}^-+\ \, \text{Cl}^-+6\text{H}^+& \rightarrow 6\text{Cl}\text{O}_{2}+3\text{H}_{2}\text{O}.
\end{align*}
These HBRs give a complete description of the solution set $S$. On one hand, any reactions from the reactants to the products can be generated by these HBRs; on the other hand, any combination of HBRs (with nonnegative integer coefficients) is a valid reaction of this system. This approach based on Hilbert basis avoids the shortcomings mentioned above for the conventional approach.


In the second example, we consider a complex reaction involved in the production of perchloric acid~\cite{mcbride_g_1984,jensen_unbalanced_1987}. Here, the reactants consist of four substances
$
  R=\{\text{N}\text{H}_{4}\text{Cl}\text{O}_{4},\text{H}\text{N}\text{O}_{3},\text{H}\text{Cl},\text{H}_{2}\text{O}\}
$, and the products contain five $P=\{\text{H}\text{Cl}\text{O}_{4}\cdot2\text{H}_2\text{O},\text{N}_{2}\text{O},\text{N}\text{O},\text{N}\text{O}_{2},\text{Cl}_{2}\}$.
In other words, we are balancing the following chemical equation:
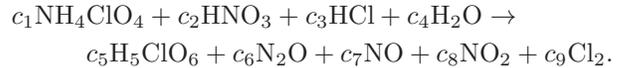
\begin{align*}
c_1 \text{N}\text{H}_{4}&\text{Cl}\text{O}_{4}+c_2\text{H}\text{N}\text{O}_{3}+c_3\text{H}\text{Cl}+c_4\text{H}_{2}\text{O}\rightarrow\\
&c_5\text{H}_{5}\text{Cl}\text{O}_{6}+c_6\text{N}_{2}\text{O}+c_7\text{N}\text{O}+c_8\text{N}\text{O}_{2}+c_9\text{Cl}_{2}.
\end{align*}
Here, we denote $\text{H}\text{Cl}\text{O}_{4}\cdot2\text{H}_2\text{O}$ as $\text{H}_{5}\text{Cl}\text{O}_{6}$ for short.
The $\mathbf F$ matrix is constructed as
\begin{equation}
\mathbf F=\left[
\begin{array}{crrrrrrrrr}
\text{Cl}&	1 & 0 & 1 & 0 & -1 & 0 & 0 & 0 & -2 \\
\text{H}&	4 & 1 & 1 & 2 & -5 & 0 & 0 & 0 & 0 \\
\text{N}&	1 & 1 & 0 & 0 & 0 & -2 & -1 & -1 & 0 \\
\text{O}&	4 & 3 & 0 & 1 & -6 & -1 & -1 & -2 & 0 \\
\end{array}
\right].
\end{equation}

The null space $\ker\mathbf F$ is five dimensional. Its five basis vectors are obtained as
\begin{equation}
\left\{
\begin{array}{lrrrrrrrrr}
 \mathbf b^1&	-6 & 6 & 13 & 0 & -1 & 0 & 0 & 0 & 4 \\
 \mathbf b^2&	-2 & 10 & 3 & 0 & 1 & 0 & 0 & 8 & 0 \\
 \mathbf b^3&	2 & 6 & 1 & 0 & 3 & 0 & 8 & 0 & 0 \\
 \mathbf b^4&	1 & 1 & 0 & 0 & 1 & 1 & 0 & 0 & 0  \\
 \mathbf b^5&	-1 & 1 & 1 & 1 & 0 & 0 & 0 & 0 & 0 \\
\end{array}
\right\}.
\end{equation}
It is noted that $\mathbf b^1, \mathbf b^2$ and $\mathbf b^5$ contain negative coefficients, so they are not valid solutions. And it is nontrivial how to make valid solution via linear combinations of these vectors.

Meanwhile, by the Hilbert basis approach, we find there are a unique set of 17 Hilbert basis vectors for this problem, which are given by
\[
\left\{
\begin{array}{lccccccccc}
	
&	
\tiny{\text{N}\text{H}_{4}\text{Cl}\text{O}_{4}}&\tiny{\text{H}\text{N}\text{O}_{3}}&\tiny{\text{H}\text{Cl}}&\tiny{\text{H}_{2}\text{O}}&
	\tiny{\text{H}_{5}\text{Cl}\text{O}_{6}}&\tiny{\text{N}_{2}\text{O}}&\tiny{\text{N}\text{O}}&\tiny{\text{N}\text{O}_{2}}&\tiny{\text{Cl}_{2}}\\
\mathbf v^{1}&0&2&1&1&1&1&0&0&0\\
\mathbf v^{2}&0&4&1&0&1&0&2&2&0\\
\mathbf v^{3}&0&6&2&1&2&0&5&1&0\\
\mathbf v^{4}&0&6&4&0&2&0&6&0&1\\
\mathbf v^{5}&0&6&4&0&2&1&3&1&1\\
\mathbf v^{6}&0&6&4&0&2&2&0&2&1\\
\mathbf v^{7}&0&8&2&0&2&1&1&5&0\\
\mathbf v^{8}&0&8&3&2&3&0&8&0&0\\
\mathbf v^{9}&0&8&7&0&3&2&4&0&2\\
\mathbf v^{10}&0&8&7&0&3&3&1&1&2\\
\mathbf v^{11}&0&10&10&0&4&4&2&0&3\\
\mathbf v^{12}&0&12&3&0&3&2&0&8&0\\
\mathbf v^{13}&0&12&13&0&5&6&0&0&4\\
\mathbf v^{14}&1&1&0&0&1&1&0&0&0\\
\mathbf v^{15}&1&5&1&0&2&0&5&1&0\\
\mathbf v^{16}&1&7&2&1&3&0&8&0&0\\
\mathbf v^{17}&2&6&1&0&3&0&8&0&0
\end{array}
\right\}.
\]
The corresponding HBRs can be readily written out. For example, the first two basis $\mathbf v^{1}$ and $\mathbf v^{2}$ lead to the following two HBRs:
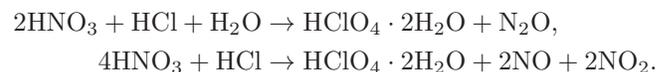
\begin{align*}
2\text{H}\text{N}\text{O}_{3}+\text{H}\text{Cl}+\text{H}_{2}\text{O}&\rightarrow \text{H}\text{Cl}\text{O}_{4}\cdot2\text{H}_2\text{O}+\text{N}_{2}\text{O},\\
4\text{H}\text{N}\text{O}_{3}+\text{H}\text{Cl}&\rightarrow \text{H}\text{Cl}\text{O}_{4}\cdot2\text{H}_2\text{O}+2\text{N}\text{O}+2\text{N}\text{O}_{2}.
\end{align*}
Any valid solution can be generated by these HBRs.

\section{conclusion}

In conclusion, we propose a new algebraic approach for balancing chemical equations. It is based on the recognition of the proper algebraic structure of solution set being positive affine monoid rather than linear space. This understanding leads to a unique set of elementary solutions, the HBRs, corresponding to the Hilbert basis of positive affine monoid. 
Our approach avoids several problems with conventional algebraic approaches, such as invalid solutions with negative ceofficients, non-uniqueness and incorrect number of independent elementary reactions. We have developed a open source package 
\textsf{HilbertBalance}, which implements our approach and can be readily used on a personal computer. 

\begin{acknowledgements}
The authors thank D. L. Deng for valuable discussions. We acknowledge support from the Ministry of Education of the People's Republic of China's Higher Education University Physics Curriculum Teaching Guidance Committee and the Transition Working Committee for Secondary and Higher Physics Education (No. WX202448). The Fundamental Research Funds for the Central Universities (No. ZY2418) and UM start-up grant (SRG2023-00057-IAPME).
\end{acknowledgements}

\bibliography{BalanceChem1}

\end{document}